\documentclass[11pt]{article}
\usepackage{geometry}                
\usepackage{makeidx}
\usepackage[pdftex]{graphicx}
\usepackage{lscape}
\usepackage{amssymb}
\usepackage{epstopdf}
\usepackage{amsmath,amssymb,amsthm}
\usepackage{latexsym}
\usepackage{listings}
\usepackage{chemarr}
\pdfoutput=1
\newtheorem{axiom}{Axiom}[section]
\newtheorem{defin}{Definition}[section]
\newtheorem{theo}{Theorem}[section]
\newtheorem{propo}{Proposition}[section]
\newtheorem{lemma}{Lemma}[section]
\newtheorem{corol}{Corollary}[section]
\newtheorem{remark}{Remark}[section]
\newcommand{\bax}{\begin{axiom}}
\newcommand{\eax}{\end{axiom}}
\newcommand{\bass}{\begin{assumption}}
\newcommand{\eass}{\end{assumption}}
\newcommand{\bdefi}{\begin{defin}}
\newcommand{\edefi}{\end{defin}}
\newcommand{\bth}{\begin{theo}}
\renewcommand{\eth}{\end{theo}}
\newcommand{\bprop}{\begin{propo}}
\newcommand{\eprop}{\end{propo}}
\newcommand{\blem}{\begin{lemma}}
\newcommand{\elem}{\end{lemma}}
\newcommand{\bcor}{\begin{corol}}
\newcommand{\ecor}{\end{corol}}
\newcommand{\brem}{\begin{remark}}
\newcommand{\erem}{\end{remark}}
\newcommand{\bpf}{\begin{proof}}
\newcommand{\epf}{\end{proof}}

\newcommand{\noi}{\noindent}
\newcommand{\bn}{{\bf n}}

\newcommand{\bk}{{\bf k}}
\newcommand{\bx}{{\bf x}}
\newcommand{\vde}{{\vec{\de}}}
\newcommand{\field}[1]{\mathbb{#1}}

\newcommand{\R}{\field{R}}

\newcommand{\N}{\field{N}}
\newcommand{\X}{\field{X}}
\newcommand{\LL}{\field{L}}
\newcommand{\pa}{\partial}

\newcommand{\ra}{\rightarrow}

\newcommand{\rlha}{\xrightleftharpoons}
\newcommand{\de}{\delta}

\newcommand{\sig}{\sigma}
\newcommand{\Sig}{\Sigma}
\newcommand{\eps}{\epsilon}
\newcommand{\veps}{\varepsilon}
\newcommand{\A}{{\mathcal A}}

\newcommand{\Eop}{{\bf E}}

\newcommand{\Lop}{{\widehat{{\mathcal L}}}}
\newcommand{\Aop}{{\widehat{{\mathcal A}}}}
\newcommand{\Pro}{{{{\mathcal P}}}}

\newcommand{\Ld}{{{{\mathcal L}}}}
\newcommand{\Kd}{{{{\mathcal K}}}}
\newcommand{\lop}{{\widehat{\ell}}}
\newcommand{\id}{{\bf id}}

\newcommand{\be}{{\bf 1}}

\newcommand{\rank}{\it{rank}}
\newcommand{\idf}{\emph{i.d.f.}}
\newcommand{\fdf}{\emph{f.d.f.}}
\newcommand{\app}{\approx}

\begin{document}

\title{Molecular Systems with Infinite and Finite Degrees of Freedom. Part II: Deterministic Dynamics and Examples}


\author{
Luca Sbano, 
Mathematics Institute, University of Warwick \\
              \texttt{sbano@maths.warwick.ac.uk}\\         
              and\\     
              Markus Kirkilionis
              Mathematics Institute, University of Warwick \\
              \texttt{mak@maths.warwick.ac.uk}  
}

\date{}

\maketitle

\begin{abstract}
In this paper we consider deterministic limits of molecular stochastic systems with finite and infinite degrees of freedom. The method to obtain the deterministic vector field is based on the continuum limit of such microscopic systems which has been derived in \cite{sbano1}. With the aid of the theory we finally develop a new approach for molecular systems that describe typical enzyme kinetics or other interactions between molecular machines like genetic elements and smaller 'communicating' molecules. In contrast to the literature on enzyme kinetics the resulting deterministic functional responses are not derived by time-scale arguments on the macroscopic level, but are a result of time scaling transition rates on the discrete microscopic level. We present several examples of common functional responses found in the literature, like Michaelis-Menten and Hill's equation. We finally give examples of  more complex but typical macro-molecular machinery.
\end{abstract}

\section{Introduction}

 In this second part we study first the deterministic limit of systems involving small 'communicating' molecules and bigger macro-molecules, which in the following we will call 'molecular machines'. The communicating molecules will be described as concentrations, with the justification that we assume large number of molecules in the reaction volume and have derived the continuum limit for this part of the system. In contrast to this, the macro-molecular machines being considered will typically only occur in finite numbers. Moreover this number will not change during the observation of the system. In case numbers are small, we expect the system will have more variance in the observables, i.e. it will be more 'noisy'.  The construction in  \cite{sbano1} did describe such a system in terms of a master equation (ME), where for any communicating molecular species with label $i$, $1 \le i \le N$, the evolution of species numbers was described by the probability  $P_\sig(\bn,t)$ given by

\begin{equation}
\frac{\pa P_\sig(\bn,t)}{\pa t}=\Ld_\sig^*(\bn)(P_\sig(\bn,t))+\sum_{\sig'\in \Sig}\Kd^T_{\sig\sig'}(\bn)\,P_{\sig'}(\bn,t)\mbox{ with $\sig\in \Sig$,}
\label{main-ME00}
\end{equation}

where $\bn=(n_1,...,n_N)\in\N^N$ represents the collection of numbers of these 'communicating' molecules of different species, $\sig \in S$ is an index for all different discrete states of the Markov chain describing the collection of all molecular machines in the system, $\Ld^*_\sig(\bn)$ is a collection of difference operators stemming from a birth-death process, and $\Kd^T(\bn)$ is the transpose of a generator of a Markov chain on $S$. 
The above system was defined in  \cite{sbano1} as a microscopic system with infinite and finite degrees of freedom, or short, an IFSS. Note that as we do not include spatial position to any entity, the discrete state space  $\Sig$ indeed describes either one or several finitely many molecular machines, all equally accessible by the 'communicating' molecules in one compartment. In other words it describes any discrete state attached to any of the macro-molecules in one ore several copies inside one such  'well-mixed' compartment. The case of several copies of identical machines is subsequently described by  having zero transition rates between analogous states. The respective structuring of $\Sig$ with this interpretation is discussed in detail in section \ref{constructionS}. The continuum limit as derived in \cite{sbano1} next  transforms the ME into a Fokker-Planck equation (FPE) of the form:
\begin{equation}
\frac{\pa\rho}{\pa t}=\Lop\,\rho+\frac{1}{\eps}\,\Kd\,\rho
\label{cont_lim}
\end{equation}
where $\eps$ is a parameter dependent on the physic scales of the system. The  equation (\ref{cont_lim}) is obtained by letting to zero the discretisation implicit in the Master Equation and has an asymptotic validity. This is called continuum limit and in its regime $\eps$ is infinitesimal. Physically this corresponds to that  the dynamics of the finite state Markov chain (finite degrees of freedom) is faster than the dynamics of transitions between the communicating smaller molecules, as described by the infinite degrees of freedom.\\ 
Upon $\eps\simeq 0$, by application of an asymptotic expansion we derived a solution of equation (\ref{cont_lim}). This was called \emph{adiabatic approximation}. The resulting leading order term constitutes a deterministic dynamics which turns out to be an average over the different invariant measures of the Markov chain.  We call this dynamics also  \emph{average dynamics}. In \cite{sbano1} it is shown that the average dynamics is given as a system of ordinary differential equations involving the concentrations, and (a convex combination of) invariant measures derived by the asymptotic limit of the Markov chain (MC). The average dynamics is given by
 
 \begin{equation}
 \frac{d x_i(t)}{dt}=\sum_{\sig\in \Sig}\mu_\sig(x_1(t),...,x_N(t))\,A_i^{(\sig)}(x_1(t),...,x_N(t)),~~~i=1,...,N,
 \end{equation}
 
\noi where $x_i$ is the concentration associated to $n_i$ and $\mu$ is an invariant measure for the MC on $S$. Note that we allow 
$\mu$ to be a convex combination of invariant measures. The functions $A_i^{(\sig)}$ are a collection vector fields describing the dynamics associated to the deterministic limit of the ME. With the help of this setting we will develop an alternative approach to the description of enzyme kinetics, genetic interactions, ion channel dynamics and other macro-molecular performances in a typical cell. As already mentioned we assume no spatial structure in this approach explicitly. Spatial relationships are encoded in the transition and reaction rates of the system, like in the case of mass-action kinetics.  Also typical enzyme kinetics are based on mass-action kinetics with an additional time scaling argument. In this respect the theory developed in  \cite{sbano1} and this second part is completely analogous. Nevertheless we will be able to derive compartmental models, and will have one example for such a system. An extension to spatially distributed concentrations for the communicating molecules appears to be very natural, keeping the discrete nature of the bigger in scale macro-molecular machines. We start developing the theory by recalling some properties of the continuum limit and the adiabatic approximation.

\section{The continuum limit and the adiabatic approximation}
Let us now recall the setting introduced in \cite{sbano1} to construct the Master Equation.
The interaction among molecules depends on the scales at which we study the specific system. In particular the interactions depend on the number of molecules involved: the size scales of the system. When few molecules interact the stochastic nature of the 
law of Physics prevails, but as the concentrations increase deterministic effects begin 
to emerge. To capture this transition it is necessary to build the description of 
the dynamics in a form that makes explicit reference to the scales.  
\bdefi
Let us define two sets of scales
\begin{enumerate}
\item size scales $\vde=(\de_1,...,\de_N)$, $\de_i>0$,
\item time scale $\tau>0$.
\end{enumerate}
The size scales $\vde$ describe the level at which the number of particles are counted. 
The time scale $\tau$ is the "time-step" at the processes (e.g. chemical reactions)
 take place. We shall see that the \emph{continuum limit} will be the formalisation of 
 the taking $\tau\ra 0$ and $\bn\ra \infty$, $\vde\ra 0$ 
 keeping $\bn\,\vde$ finite.\\
 The scales $\vde$ are used to define the space where the chemical processes
  take place:\\ 
Let $\LL_\vde$ be the following lattice
\begin{equation}
\label{lattice}
\LL_\vde\doteq\{\bn\,\vde=(n_1\,\de_1,...,n_N\,\de_N):~~\bn=(n_1,...,n_N)\in\N^N\}
\end{equation}
\edefi
\brem
On $\LL_\vde$ we shall define functions, now for fixed $\vde$ the value of any function $u$ 
is uniquely determined by the integer vector $\bn$  therefore whenever $\vde$ is fixed we shall omit the $\vde$ dependence and write $u(\bn)$.
\erem
At a fixed time the state of a reaction is system is clearly defined by the number 
of different type of particles and this corresponds to a point in $\LL_\vde$. Species of particles that can be in any number form the \emph{infinite degrees of freedom (i.d.f.)} 
of the given system.\\
In many reaction network, in particular in biological  systems, there degrees of freedom 
that cannot be described as points in some lattice $\LL_\vde$. Indeed conformational changes in molecules and binding/unbinding events are typical example of 
configurations which are discrete and finite in nature. To describe such digrees of freedom that we called \emph{finite degrees of freedom (f.d.f.)} we introduced 
a finite set of symbols denoted by $\Sig$. Therefore we define
\bdefi
The state $\zeta$ of the system is fully specified by $n_1,...,n_N$ \emph{infinite degrees of freedom} (\emph{i.d.f.}) and a second variable, the \emph{finite degrees of freedom} $s$ (\emph{f.d.f.}). The state $\zeta$ is therefore the composition

\[\zeta=(n_1\de_1,....,n_N\de_N,\sig)=(\bn\vde,\sig) \in \LL_\vde\times \Sig,\]

where $\LL_\vde=\vde\N^N$, $\bn$ is an $n$-tuple of natural numbers and $\sig$ runs in a finite set $\Sig$, with $|\Sig|=g$ being the number of discrete states. 
\edefi

Now the time evolution is determined by a stochastic dynamics, this motivates the following definition: 
\bdefi
Let the  tuple $(\zeta, R, P)$ determine a stochastic process by specifying the state $\zeta$, a set of reactions $R$, and a vector of probabilities $P$, such that

\begin{itemize}
\item[(i)] $\zeta$ valued in $\LL_\vde\times\Sig$, 

\item[(ii)] the time evolution of the stochastic process is defined via the set of reactions $R$ having three different types:

\begin{enumerate}
\item[(a)] Processes involving only \idf's represented by reactions (possibly reversible) of the form

\[(\bn,\sig)\ra(\bn',\sig).\] 

The operator describing these reactions in the master equation  is denoted by $\Ld^*_R$ and has the form
$\Ld^*_R=\ell_0\otimes\delta_{\sig\sig'j}$ where $\ell_0$ is the same operator for each discrete state $\sig=1,...,g$. Here $\delta_{\sig\sig'} =1$ for $\sig=\sig'$, and zero otherwise.

\item[(b)] Processes involving only \fdf's represented by 
reactions (possibly reversible) of the form

\[(\bn,\sig)\ra(\bn,\sig').\] 

The operator describing these reactions in the master equation  is  the transpose $\Kd^T$ of  the Markov chain generator of  the process governing the transitions among the discrete states $\sig=1,...,g$.  The Markov chain is finite dimensional with a space of stationary states $M_\Kd$ of dimension strictly less than g.
 
\item[(c)] Processes involving both \idf~ and \fdf~  represented by 
reactions (possibly reversible) of the form

\[(\bn,\sig)\ra(\bn',\sig).\] 

The operator describing these reactions in the master equation is denoted by $\Ld^*_E$. This operator is non-trivial only in the discrete states $\sig$ which affect processes involving \idf.

\end{enumerate}
\end{itemize}

\item[(iii)] each realisation of the process is valued in $\LL_\vde^N\times \Sig$. The state $\zeta$ at time $t$ is given by the vector of probabilities

\begin{equation}
P(t,\bn)=(P_1(t,\bn),...,P_g(t,\bn)), \mbox{ with } \sum_{\bn\in\N^N}\sum_{\sig=1}^gP_\sig(t,\bn)=1. 
\label{e:problem}
\end{equation}

The time evolution of $P$ is given by the master equation (ME)

 \begin{equation}
 \frac{\pa P(t,\bn)}{\pa t}=(\Ld^*_R+\Ld^*_E)\circ P(t,\bn)+ \Kd^T(\bn)\,P(t,\bn), 
 \label{equation0}
 \end{equation}

$P$, $\Ld^*_R$, $\Ld^*_E$ and $\Kd^T$ are sufficiently regular such that (\ref{equation0}) has a unique solution for all times $t >0$. Then the tuple $(\zeta, R, P)$ is called a (microscopic) system with \emph{infinite and finite degrees of freedom}, or short an $\bf IFSS$ (Infinite-Finite State System).
\edefi

\subsection{Construction of the continuum approximation}
\label{continuum}
The ME results from the specification of the reactions at a given scales $\vde,\tau$. 
Equation (\ref{equation0}) describes the evolution in time of the probability distribution 
$P$. To understand the technical structure of the continuum limit one needs to make the following preliminary observation. A probability being a measure  can be a very "non-smooth" object and therefore in the limit $\vde\ra 0$, $\tau \ra 0$ the function $P$ 
is not expected to have in general a "smooth" limit.\\
To overcame this difficulty the idea (see \cite{Kurtz}) is to look for a limit in the 
space of function which are dual to the space of probability  measures. The dual of the space of measures is in fact a more tame object, and in such a space the ME (\ref{equation0}) has an \emph{adjoint} formulation, for which the limit can be formulated (see \cite{Kurtz,Pazy,sbano1}).\\
The ME describes the evolution of a probability measure $P_\sig(t;\bn)$ according to
\begin{equation}
\frac{\pa P}{\pa t}=\A^*[\vde,\tau]P,
\label{eq-adjA}
\end{equation}
where $\A^*[\vde,\tau]$ is the infinitesimal generator defined on the scales $\vde,\tau$ by
\begin{equation}
\A^*[\vde,\tau]\doteq\Ld^*[\vde,\tau]+\Kd^T[\vde,\tau].
\label{op-A}
\end{equation}
The operator $\A^*[\vde,\tau]$ is defined on the space 
\begin{equation}
\X^*_{\vde,\tau}\doteq\left\{P_\sig(t;\bn):\sum_{\bn\in\LL_\de}\sum_{\sigma\in\Sigma}
P_\sig(t;\bn)=1\mbox{  for al $t$}\right\}.
\label{dualXn}
\end{equation}
Let us now consider a sequence of scales $\vde_n,\tau_n$ such that $\vde_n\ra 0$ 
and $\tau_n\ra 0$ as $n\ra \infty$. For each index $n$ we have an operator $\A_n=\A[\vde_n,\tau_n]$ defined on $\X^*_n=\X^*_{\vde_n,\tau_n}$ where the configuration space can now be denoted by $\LL_n=\LL_{\vde_n}$. We ask ourselves what would the fate of (\ref{op-A}) be as $n\ra\infty$.\\
We can think of $\vde_n\ra 0$ and $\tau_n\ra 0$  as limit at which space and time step 
become continuous and the numbers of particles are sufficiently to be accounted as densities and this motivates the name \emph{continuum limit}.\\
The formulation of the continuum limit can be obtained  by using the approximation 
scheme introduced by Trotter in \cite{Trotter}, (see also \cite{Pazy}, \cite{Kurtz}). 
As in \cite{sbano1} to each $\A^*_n$ defined on $\X^*_n$ we can associate a vector space $\X_n$ and an adjoint operator $\A_n$.
The vector space is defined by
\begin{equation}
\label{Xspace}
\X_{n}=\X_{\vde_n,\tau_n}\doteq\left\{u(t;\bn,\sigma):\LL_{n}\times\Sigma\ra\R^g:
\|u\|_\infty=\sup_{\bn\in\LL_\de,\sigma\in\Sigma}|u_\sig(t,\bn)|<\infty, \mbox{  for al $t$}\right\}.
\end{equation}
Each $\X_n$ is dual to $\X^*_n$ according to the pairing:
\begin{equation}
\label{nduality}
\langle u,P\rangle_n\doteq\sum_{(\bn,\sigma)\in\LL_n\times\Sigma}u_\sig(\bn)\,P_\sig(\bn).
\end{equation}
The adjoint $\A_n$ is defined by:
\begin{equation}
\langle\A_nu,P\rangle_n=\langle u,\A^*_nP\rangle_n.
\end{equation}
Let us consider
\[u(t,\bn,\sig)=\sum_{\bn',\sig'}P(t,\bn,\bn',\sig,\sig')u(\bn',\sig')\]
then to equation (\ref{eq-adjA}) we now associate
 \begin{equation}
\frac{\pa u}{\pa t}=\A_n\,u,
\label{eq-A}
\end{equation}
defined on each $\X_n$. Here
\begin{equation}
\A_n\doteq\lim_{t\ra 0}\frac{1}{t}(P^t-\id)
\end{equation}
see \cite{Wentzell} for all the details.\\
 For index $n$ equation (\ref{eq-A}) is the standard 
Kolmogorov and $\A_n$ is the infinitesimal generator of Markov process 
on $\LL_n\times\Sigma$.\\
The definition of the continuum limit is based on the choice of a target space 
where the limit is attained. We shall consider as target the space of 
continuous function $\X=C^0(\R^N,\R^g)$.  The topological dual of $\X$ is formed 
by signed measures on $\R^N\times\Sigma$:
\begin{equation}
\X^*=\left\{\rho(\bx):\langle\rho,u\rangle<\infty,~~u\in\X\right\},
\end{equation}
where the pairing is defined by 
\[\langle\rho,u\rangle\doteq \int_{\R_+^N}d\bx\,\sum_{\sigma\in\Sigma}\rho_\sigma(\bx)
u_\sigma(\bx).\]
According to \cite{Trotter} we define a sequence of projections
\bdefi
Let $\Pro_n:\X\mapsto\X_n$ be the operator that maps $u\in\X$ to
$\Pro_n(u)\in\X_n$ defined as
\[\Pro_n(u)(\bk)=u(\bk\,\vde_n)=u(k_1\de^1_n,...,k_N\de_n^N).\]
\edefi
The following holds true (see \cite{sbano1})
\bprop
The projections $\Pro_n$ satisfy the following properties
\begin{itemize}
\item[(i)] $\|\Pro_n\|_n<1$,
\item[(ii)] $\lim_{n\ra\infty}\|\Pro_n(u)\|_n=\|u\|_\infty$ for every $u\in\X$.
\end{itemize}
\eprop
Following (\cite{Trotter}) the projectors $\Pro_n$ allow to define in what sense 
the spaces $\X_n$ approximate $\X$.

\bdefi
A sequence $u_n\in\X_n$ converges to $u\in\X$ if 
\[\|\Pro_n(u)-u_n\|_n\ra 0\mbox{ as $n\ra\infty$}.\]
We denote this by $u_n\app u$. 
\edefi
We now give the definition for the limit, in fact the \emph{continuum limit}, 
of a sequence of operators $\A_n$. This definition is inspired by the one presented in \cite{Trotter}. In fact in the present case we have to consider that the operators are functions of the scales $\vde_n$ and $\tau_n$. Therefore we set
\bdefi
Let $\A_n:\X_n\mapsto\X_n$ be a sequence of linear operators. We say that 
$\Aop:\X\mapsto\X$ is the continuum limit of $\A_n$ (denoted by $\A_n\app\Aop$) if there exists a sequence of scales $\vde_n,\tau_n$ such that
\begin{enumerate}
\item $\vde_n\ra 0$, $\tau_n\ra 0$,
\item the domain of $\A$ is 
\[D(\Aop)=\{u\in\X:~\Pro_n(u)\in D(\A_n),~~ \A_n(\Pro_n(u))\mbox{ converges}\},\]
\item and $\|\Pro_n(\Aop(u))-\A_n(\Pro_n(u))\|_n\ra 0$ as $n\ra\infty$.
\end{enumerate}
\edefi

\brem
The dependence on the choice of the scales $\vde_n$ and $\tau_n$ makes the continuum limit non unique.  This is very important because with the choice of the scaling 
we will be able to analyse different type of processes.  
\erem

Let us now recall two examples of continuum limits that will be used in what follows.\\
We consider
\begin{equation}
\A^*=\frac{1}{\tau}(\Eop^+-\id)
\label{first-examp}
\end{equation}
defined on $\X_{\de,\tau}^*$ as follows:
\[\A^*(P)(m)=\frac{1}{\tau}(\Eop^+-\id)P(m)=\frac{P(m+1)-P(m)}{\tau}.\]
The adjoint $\A$ defined on $\X_n$ is
\[\A(u)(m)=\frac{1}{\tau}(\Eop^--\id)u(m)=\frac{u(m-1)-u(m)}{\tau},\]
the continuum limit computed in \cite{sbano1} is
\[\A\app\Aop=-c\frac{\pa}{\pa x}\]
for $\de\ra 0,\tau\ra 0$ with $\de/\tau=c>0$. $\Aop$ is densely defined on the space of continuous functions. The operator $\Aop$ has an adjoint on the space of measures $\rho$ defined by
\[\Aop(\rho)=\frac{\pa}{\pa x}(c\,\rho).\]

Now let us consider the matrix operator
\begin{equation}
\Kd^T=\frac{1}{\tau}\left(\begin{array}{cc}
-m\,k^+(\de,\tau) & k^-(\de,\tau)\\
m\,k^+(\de,\tau) & -k^-(\de,\tau)\\
\end{array}\right).
\label{second-examp}
\end{equation}
Its adjoint is
\[
\Kd=\frac{1}{\tau}\left(\begin{array}{cc}
-m\,k^+(\de,\tau) & m\,k^+(\de,\tau)\\
k^-(\de,\tau) & -k^-(\de,\tau)\\
\end{array}\right).\]
under the condition that for $\de\ra 0, \tau\ra 0$ and $\eps=\eps(\de,\tau)=\tau$ such that
\[\frac{k^+(\de,\tau)}{\tau}\simeq \frac{\de\,k^+}{\eps},~~
\frac{k^-(\de,\tau)}{\tau}\simeq \frac{k^-}{\eps}\]
then continuum limit  is 
\[\Kd=\frac{1}{\eps}\left(\begin{array}{cc}
-x\,k^+& x\,k^+\\
k^- & -k^-\\
\end{array}\right).\]
\brem
It is important to observe in the study of the ME the limits $\de\ra0$, $\tau\ra 0$ involve 
two choices that have to be compatible:
\begin{itemize}
\item[(i)] the limit of the ratio $\de^p/\tau$ for som $p>0$,
\item[(ii)] the choice of the $\eps(\de,\tau)$.
\end{itemize}
\erem
In the previous example the important assumption is to have 
\[k^+(\de,\tau)=\de\,k^+,~~~k^-(\de,\tau)=k^-.\]
It is sufficient that these conditions are satisfied asymptotical as $\de\ra 0,~\tau\ra 0$.

\subsection{Average deterministic dynamics}
Generalising \cite{kepler-elston} we assume that the Markov chain has possibly more than one stationary measure

\[M_K\doteq\{\mu(\bx):\Kd^T(\bx)\mu(\bx)=0\}.\]

From a modelling point-of-view we assume that we either have a weighted average of residence times in which the Markov chain resides in equilibrium, implicitly assuming an additional 'microscopic noise' not explicitly modelled which triggers a transition from one invariant measure to the next with fixed rates. Or we assume to have a population of identical machines like enzymes in which a fraction is in one feasible equilibrium, another fraction of the population in the next equilibrium etc.. We avoid trivialities by assuming $m_\Kd\doteq\dim(M_\Kd)<g.$ Any convex combination

\[\mu=\sum_{m=1}^{m_\Kd}\theta_m\,\mu^{(m)}\mbox{ with } \sum_{m=1}^{m_\Kd}\theta_m=1\]

\noi is in $M_\Kd$ (see \cite{Brzezniak}). Note that $M_\Kd$ correspond also to all possible invariant measures of the Markov chain. Each such measure describes the possible asymptotic behaviour of the Markov chain which is in general decomposable, i.e. a product of $m_\Kd$ Markov chains. We now take one convex combination $\mu\in M_\Kd$ and construct the adiabatic theory for the FPE obtaining an asymptotic expansion in $\eps$ of $\rho$. Let us next consider the marginal distribution defined by

\begin{equation}
f(\bx,t)=\sum_{\sig\in \Sig_\mu}\rho_\sig(\bx,t)\mbox{ with } \Sig_\mu=\{\sig\in \Sig:\mu_{\sig}\neq 0\}.
\end{equation}

\noi In \cite{sbano1} we showed that  $f(\bx,t)$ can be expanded in $\eps$ and made accurate up to order $O(\eps)$. Then $f(\bx,t)$ is determined by

\begin{equation}
\frac{\pa f(\bx,t)}{\pa t}=\sum_{m=1}^{m_\Kd}\sum_{\sig\in \Sig}\theta_m\Lop_\sig^*(\bx)(\mu^{(m)}_\sig(\bx)f(\bx,t))+\eps\Gamma(\bx)[f(\bx,t)],
\label{average-dyn0}
\end{equation}

\noi where $\Gamma$ is a parabolic operator. In particular 

\[\Gamma(x)[f]=\sum_{\alpha\beta}\frac{\pa^2}{\pa x_\alpha\pa x_\beta}(\Gamma(\bx)f(\bx,t)), \]

\noi where

\[\Gamma_{\alpha\beta}(\bx)=-\sum_{j,m}L_\beta^m(\bx)(\Kd^T)^D_{mj}(\bx)L_\alpha^j(\bx)\mu_j(\bx),
\]

\noi and $(\Kd^T)^D(\bx)$ is the so called Drazin inverse of $\Kd^T(\bx)$. 
 Now one can observe that  the operator $\Gamma$ is parabolic leading to the applicability of standard results from probability theory (see  \cite{gardiner,oksendal}). Such results guarantee that a solution $f(\bx,t)$ of (\ref{average-dyn0}) is the probability distribution of a Markov process. In the next section we recall the link between $f(\bx,t)$  and the time evolution of the concentrations $\bx(t)$.

\subsection{Description in terms of SDEs and ODEs}

The FPE (\ref{average-dyn0}) describes the time evolution (up to order of the $O(\eps)$) of the probability density $f$ of a Markov process. A standard result 
in probability theory (see again \cite{gardiner,oksendal}) links the FPE to an Ito stochastic differential equation (SDE) which gives the trajectory, i.e.  the realisations of the Markov process. One can show that in our case the  stochastic differential equation defined on a finite time interval is 

\begin{equation}
dx_\alpha(t)=A_\alpha(\bx(t))\,dt+\sqrt{\veps}\sum_\beta\sigma_{\alpha\beta}(\bx(t))\,dw^\beta_t\mbox{, with $\alpha=1,...,N$,}
\label{SDE}
\end{equation}

and $w_t$ being an $N$-dimensional Wiener process. It holds that the noise strength satisfies $\|\sigma(\veps,\bx)\|\sim\sqrt{\eps}$, and $A(\bx)$ is the \emph{averaged vector field} given by

\begin{equation}
A_\alpha(\bx)=\sum_{\sig\in \Sig}\theta_m\,L^{\sig}_\alpha(\bx)\mu^{(m)}_\sig(\bx).
\label{average}
\end{equation}

\noi Here $L^\sig(\bx)$ is the deterministic vector field associated to the finite state $j$ and $A(x)$ is the average over the invariant measure $\mu(\bx)$ of all vector fields associated to the finite states in $\Sig$. The limit $\eps\ra 0$ taken in (\ref{SDE}) leads to the set of 
ordinary differential equations (ODE)

\begin{equation}
\frac{dx_\alpha(t)}{dt}=A_\alpha(\bx(t))\mbox{, with $\alpha=1,...,N$.}
\label{ODE}
\end{equation}

\noi We shall call (\ref{average}) and (\ref{ODE}) the \emph{average dynamics}. If $m_\Kd>1$, then the Markov chain is equivalent  to a product of $m_\Kd$ Markov chains and the vector-field  (\ref{average}) describes the deterministic dynamics averaged over all $m_\Kd$ components of $\Sig$. We illustrate the theory  using equation (\ref{ODE}) and derive as applications effective reaction rates related to different macro-molecular machinery. Prominent examples will be enzyme kinetics like Michaelis-Menten or more general, Hill's type kinetics. In a forthcoming paper we apply this theory to derive rigourously the nonlinear macroscopic model used in \cite{thattai} to study - on a more heuristic basis - the bistability of the \emph{Lac-Operon} switch inside this framework.

\section{Explicit construction of the discrete state space $\Sig$ and the average dynamics}
\label{constructionS}

In this section we analyse the construction of the space $\Sig$ and the consequences on the average dynamics. This will help in the understanding of the theory in relation to applications, like the derivation of enzyme kinetics. 

\subsection{The space $\Sig$}

In modelling it is crucial to construct the discrete state space $\Sig$ of the Markov chain in a meaningful and consistent way. For example the modelling step will involve  the identication of different types of interacting molecular machines present in a process., Each such machine will have its own set of different states, its finite degrees of freedom forming a subset of $\Sig$. We therefore give an explicit construction of $S$ in terms of compositions of subspaces structuring $\Sig$. Let us consider a system formed by a certain number of chemical species $\bn=(n_1,...,n_N)$  that can take any integer value, i.e. these small molecules can be present in any number in the system. This means $\bn\in\N^N$, $\bn$ is an \idf~ and will be treated as a birth-death process, see \cite{sbano1}. Furthermore let us suppose there are now $M < \infty$ macro-molecules $(\sig_1,...,\sig_M)$, each of which  can take only a finite number of conformations, i.e. forming subset of $\Sig$. It is useful to introduce the notation

\[\sig_{ij}\in \Sig_i\mbox{, with $i=1,...,M$,}\]

\noi where $\Sig_i=\{\sig_{i1},...,\sig_{ig_i}\}$. Each $\Sig_i$ is the finite set of all possible states for $\sig_{ij}$ with $|\Sig_i|=g_i$. We have two possible ways to construct the total space $S$ out of these sub-spaces:

\bdefi[Product Space]
The \fdf~ space consisting of all possible conformations $(\sig_1,...,\sig_M)$ is given by the Cartesian product

\[\Sig=\times_{i=1}^M\Sig_i,\]

\noi where each $\Sig_i$ is finite and therefore $|\Sig|=\Pi_{i=1}^Mg_i=g$.\\
\edefi

\bdefi[Direct Sum Space]
The \fdf~ space consisting of all possible conformations $(\sig_1,...,\sig_M)$ is given by the Cartesian product

\[\Sig=\oplus_{i=1}^M\Sig_i,\]

\noi where each $S_i$ is finite and therefore $|\Sig|=\sum_{i=1}^Mg_i=g$.\\
\edefi

In the construction of the ME only the transition rates in $\Sig$ enter, obviously affecting $\Kd^T$. It is therefore very useful to chose an enumeration for the elements of $\Sig$, i.e. we can write

\[\Sig=\{O_1,...,O_g\}, \]

\noi where $O_i=(\sig_{1i_1},...,\sig_{Mi_M})$, $1 \le i \le g$. We give a simple example. Suppose that we have $\Sig_1$ and $\Sig_2$ given by
 
 \[      \Sig_1=\{\sig_{11},\sig_{12}\},~~ \Sig_2=\{\sig_{21},\sig_{22},\sig_{23}\},     \]
 
\noi then the product space $\Sig=\Sig_1\times \Sig_2$ is formed by the following $6$ couples:

 \[
 \Sig=\Sig_1\times \Sig_2=\{(\sig_{11},\sig_{21}),(\sig_{11},\sig_{22}),(\sig_{11},\sig_{23}),
 (\sig_{12},\sig_{21}),(\sig_{12},\sig_{22}),(\sig_{12},\sig_{23})\}. \]
 
 \noi The direct sum $S=S_1\oplus S_2$ is formed by 
 
 \[\Sig=\Sig_1\oplus \Sig_2=\{\sig_{11},\sig_{12},\sig_{21},\sig_{22},\sig_{23}\}.\]

\subsection{Transition rates and infinitesimal generators}

As we have seen the modelling step leads naturally to the idea of  composing state spaces of Markov chains.  The corresponding infinitesimal generators related to each subspace will form equivalently the total infinitesimal generator of the system. We discuss this construction again for the direct sum and Cartesian product of  sub-spaces. It is sufficient to consider the case of two such sub-spaces, the construction can then be iterated.

\bdefi
Consider two Markov chains with state spaces $\Sig_1$, $\Sig_2$, and infinitesimal generators $\Kd_1$ and $\Kd_2$, respectively. Let 

\[\sig_{\alpha,i}\rlha[k_{\alpha,ji}]{k_{\alpha,ij}}\sig_{\alpha,j}\mbox{, for $\alpha=1,2$},\]

\noi be the transition rates between states $\sig_{\alpha,i}$ and $\sig_{\alpha,j}$ both associated to $\Sig_\alpha$. Then the transition rates in the direct sum chain $\Sig=\Sig_1\oplus \Sig_2$ are denoted by 

\[s_{\alpha,i}\rlha[k_{ji}]{k_{ij}}s_{\alpha,j}, \]

where

\[k_{ij}=\left\{\begin{array}{ll}
k_{1,ij}\mbox{ if $\alpha=1$},\\
k_{2,ij}\mbox{ if $\alpha=2$}.
\end{array}\right.\]


\edefi

\noi With the help of this notation we can make a definition on the direct sum of infinitesimal generators.
 
\bdefi
For two given Markov chains with state spaces $\Sig_1$, $\Sig_2$ and infinitesimal generators $\Kd_1$ and $\Kd_2$, respectively,  the infinitesimal generator $\Kd$ of the direct sum chain $\Sig=\Sig_1\oplus \Sig_2 $ is defined by

\[\Kd=\left(\begin{array}{cc}
\Kd_1 & 0\\
0 &\Kd_2
\end{array}\right)\]

\edefi

\noi For the product case the structure of $K$ is more complicated. In fact we have
 
\bdefi
Consider two Markov chains with state spaces $\Sig_1$, $\Sig_2$ and infinitesimal generators $\Kd_1$ and $\Kd_2$, respectively. Let 

\[\sig_{\alpha,i}\rlha[k_{\alpha,ji}]{k_{\alpha,ij}}\sig_{\alpha,j}\mbox{, for $\alpha=1,2$},\]

\noi be the transition rates  between states $\sig_{\alpha,i}$ and $\sig_{\alpha,j}$ both associated to $S_\alpha$. Then the transitions in the product chain $\Sig=\Sig_1\times \Sig_2=\{(\sig_{1i},\sig_{2j})\}_{i,j}$ are denoted by 

\[ (\sig_{1i},\sig_{2j})\rlha[k_{i'j';ij}]{k_{ij;i'j'}}(\sig_{1i'},\sig_{2j'}), \]

where

\begin{equation}
k_{ij;i'j'}=\left\{\begin{array}{lll}
k_{1,ii'}\mbox{ if $j=j'$},\\[3mm]
k_{2,jj'}\mbox{ if $i=i'$},\\[3mm]
0\mbox{ if $i\neq i'$ and $j\neq j'$}
\end{array}\right.
\label{rates-for-product}
\end{equation}

\edefi

\begin{remark}
The definition (\ref{rates-for-product}) is based on the fact that in most of  
physical examples the \emph{double} transition
\[(ii')\ra(jj')\mbox{ with $i\neq i'$ and $j\neq j'$}\]
 can be neglected.
\end{remark}

\bdefi
Let there be two Markov chains with state spaces $\Sig_1$, $\Sig_2$ respectively, and infinitesimal generators $\Kd_1$ and $\Kd_2$. Then the infinitesimal generator $\Kd$ of $\Sig=\Sig_1\times \Sig_2$ is defined by

\[K_{(1i,2j),(1i',2j')}=\left\{\begin{array}{ll}
k_{ii';jj'},\\[3mm]
-\sum_{(i',j')\in E_{ij}}k_{ii';jj'},
\end{array}\right.\]

\noi where

\[ E_{i'j'}=\{(i',j')\mbox{ is such that $(1i,2j)\ra^{k_{ii';jj'}}(1i',2j')$}\}. \]
\edefi

\noi The choice of either direct sum or product spaces as the collection of all states of $M$ distinct macro-molecular machines is therefore crucially dependent on the interpretation of the system under consideration. The product space must be used to model a situation where discrete states of different machines are coupled and cannot be attained independently, whereas the direct sum models assume complete independence of all the states associated to different machines. The notion employed to discuss the asymptotic behaviour of a Markov chain is \emph{ergodicity}. Without going into the details  a Markov chain is ergodic if each of its states during time evolution is visited again with probility $1$ and the corresponding attractor is not periodic. Ergodicity is equivalent to the existence of a unique invariant measure (see \cite{Brzezniak}). This is also equivalent to say that  the infinitesimal generator $K$ has a unique left-eigenvector (see \cite{pavliotis}).   The construction of sum and product state space structures  with ergodic Markov chains defined on the sub-spaces leads to either ergodic or non-ergodic Markov chains defined on $\Sig$, revealing the nature of each type of composition.
 
 \bprop
 Assume a collection of $M$ finite Markov chains with state spaces $\{\Sig_i\}_{i=1}^M$, where each chain 
is ergodic. Then the direct sum Markov chain is not ergodic. 
 \eprop
 
  \bpf
The direct sum of a Markov chain $\Sig$ has an infinitesimal generator $\Kd$ which is the direct sum of the generators $\Kd_i$, $i=1,\ldots,M$. Each chain $\Kd_i$ has a unique invariant measure. This implies that $\Kd$ has $M$ invariant measures and therefore $\Sig$ is not ergodic.
  \epf

 \noi We now consider a simple two-state setting to show that in contrast to direct sum state spaces the product of ergodic Markov chains  is always ergodic. This setting will be generalised in a forthcoming paper based on network theory.
    
 \bprop
Assume a collection of $M$ two-state Markov chains with state spaces $\{\Sig_i\}_{i=1}^M$ and 
generators

\[\Kd_i=\left(\begin{array}{cc}
-k_i & k_i\\
h_i & -h_i
\end{array}\right).\]

\noi If for any given $O_i,O_j\in\Sig$ there exists a sequence of rates $K_{ii_1},...,K_{i_nj}$ different from zero that allow the trasition from $O_i$ to $O_j$,  then the product Markov chain is ergodic.  
 \eprop
 \bpf
 To illustrate the situation one can consider the state space $\Sig=\{O_1,...,O_{2^M})$ with
 
 \[O_1\rlha[h_1]{k_1} O_2\rlha[h_2]{k_2}...\rlha[h_{2^M-2}]{k_{2^{M}-2}}O_{2^M-1}\rlha[h_{2^M-1}]{k_{2^{M}-1}}O_{2^M}.\]
 
 \noi This implies that the infinitesimal generator associated to the product Markov chain is such that
 
 \[\Kd_{ij}>0\mbox{, for $j=i\pm 1$, and } \Kd_{ii}<0.\] 
 
 Let $c_j=(\Kd_{ij})_{i=1}^M$ be the $j$th column vector of $\Kd$. If for each row $i$ there is $j$ such that $\Kd_{ij}>0$, then there is no columns has zero entries.  In fact 
all the diagonal entries are different from zero in $K$. Now note that
\[\sum_{j=1}^Mc_j=\sum_{j=1}^M\Kd_{ij}=\Kd_{ii}+\sum_{j\neq i}\Kd_{ij}=0\]
therefore the columns are linearly dependent and $\det(K)=0$ and thus
$\rank(\Kd)<M$.\\
Now we show that $\rank(\Kd)\geq M-1$. Assume that there exists $\lambda\in \R$ 
such that at least two columns are linearly independent:
\[c_l=\lambda\, c_m.\]
In particular this implies 
\[\Kd_{ll}=\lambda\,\Kd_{lm},~~\Kd_{ql}=\lambda\,\Kd_{qm}\mbox{ with $q,m\neq l$ and different from the possible 
row of zeros.}\]
Now using the expression for $\Kd_{ll}$ we obtain:
\[-\sum_{p\neq l,m}\Kd_{lp}=(\lambda+1)\Kd_{lm}
\mbox{ and } \Kd_{ql}=\lambda\,\Kd_{qm}.
\]
This leads to a contradiction, in fact  the parameter $\lambda$ cannot be determined since $\Kd_{ij}> 0$ for some $i,j$.  This implies $\rank(\Kd)=M-1$ and therefore 
the equation
\[\Kd^T\,\mu=\mu\,\Kd=0\]
has a unique solution.
 
 
 
 
 
\epf

 \subsection{The average dynamics and the choice of $\Sig$}
 
 The average dynamics is constructed out of two essential data:
 
 \begin{enumerate}
 \item A Markov chain with state space $\Sig$, possibly structured in finite numbers of molecular machines as discussed before.
 \item A collection of vector fields $A^{(\sig)}(\bx)$ with $\sig\in \Sig$, the different species of small molecules that will be described by concentrations.  
 \end{enumerate}
 
 \noi From $\Sig$ one can compute a stationary measure $\mu(\bx)$ and the average dynamics reads
 
 \begin{equation}
 \label{av-dyn}
 \dot{\bx}(t)=\sum_{\sig\in \Sig}\mu_\sig(\bx)\,A^{(\sig)}(\bx).
 \end{equation}
  
 \noi As just discussed in many examples the total space $S$ often results from a combination of many elementary spaces. We like to clarify the consequences of taking the total space as a product or as a sum of the elementary discrete states in more detail. Consider a  system depending on two 2-state MCs,
 
 \[\Sig_1=\{\sig_1,\sig_2\},~~\Sig_2=\{\sig_3,\sig_4\}.\]
  
 \noi In order to be able to derive the average dynamics we need to specify vector fields $A^{(\sig)}(\bx)$. The crucial point is that $A^{(\sig)}(\bx)$ may depend in many ways on the state $s$, but typically there are only two distinct options:
 
 \begin{itemize}
 \item[(i)]  $A^{(\sig)}(\bx)$ depends on each single $\{\sig_1,\sig_2,\sig_3,\sig_4\}$
  \item[(ii)] $A^{(\sig)}(\bx)$ depends on each couple $(\sig_i,\sig_j)$ with $i\neq j$.    
 \end{itemize}

\noi We examine both possibilities. The two available total spaces are
 
 \[\Sig_{sum}=\Sig_1\oplus \Sig_2=\{\sig_1,\sig_2,\sig_3,\sig_4\}\]

 \noi and

 \[\Sig_{prod}=\Sig_1\times \Sig_2=\{(\sig_1,\sig_3),(\sig_1,\sig_4),(\sig_2,\sig_4),(\sig_2,\sig_3)\}. \]

 \noi First one can note that by accident  $|\Sig_{prod}|=|\Sig_{sum}|$. Each factor chain $\Sig_i$ has the rates
 
 \[ \sig_1\rlha[h_1]{k_1}\sig_2,~~\sig_3\rlha[h_2]{k_2}\sig_4. \]
 
 \noi Using above rates one can construct the infinitesimal generator associated to $S_{sum}$. An easy calculation gives
 
 \[\Kd_{sum}=\left(\begin{array}{cccc}
 -k_1 & k_1 & 0 & 0 \\
 h_1  & -h_1 & 0 & 0\\
 0 & 0 & -k_2 & k_2\\
 0 & 0 & h_2 &-h_2
 \end{array}\right) \]
 
 \noi For $\Sig_{prod}$ we first enumerate the states
 
 \[O_1=(\sig_1,\sig_3),~~O_2=(\sig_1,\sig_4),~~O_3=(\sig_2,\sig_3),~~O_4=(\sig_2,\sig_4)\]
 
 \noi and  construct the rates
 
 \[O_1\rlha[h_2]{k_2}O_2\rlha[h_1]{k_1}O_4\rlha[h_2]{k_2}O_3\rlha[h_1]{k_1}O_1,\]
 
 \noi which allow us to write the associated generator
 
 \[
 \Kd_{prod}=\left(\begin{array}{cccc}
 -k_1 -k_2 & k_2 & k_1 & 0 \\
 h_2  & -h_2-k_1 & 0 & k_1\\
 h_1 & 0 & -k_2-h_1 & k_2\\
 0 & h_1 & h_2 &-h_2-h_1
 \end{array}\right).
 \] 
  
 \noi Now $\Kd_{sum}$ is not ergodic. In fact the equation $\Kd^T_{sum}\,\mu=0$ has two solutions
 
 \[\mu_{sum}^{(1)}=\left(\frac{h_1}{k_1+h_1},\frac{k_1}{k_1+h_1},0,0\right),\]
 
 and
 
 \[\mu_{sum}^{(2)}=\left(0,0,\frac{h_2}{k_2+h_2},\frac{k_2}{k_2+h_2}\right).\]
 
 \noi The product chain is ergodic. Using $\Kd_{prod}$ one finds the unique invariant measure 
 
 \[\begin{array}{ll}
 \displaystyle \mu_{prod}=\left({\frac {h_{{1}}h_{{2}}}{k_{{1}}h_{{2}}+h_{{1}}h_{{2}}+k_{{2}}h_{{1}}+
k_{{1}}k_{{2}}}},{\frac {k_{{2}}h_{{1}}}{k_{{1}}h_{{2}}+h_{{1}}h_{{2}}
+k_{{2}}h_{{1}}+k_{{1}}k_{{2}}}},\right.  \\[4mm]
\displaystyle \left.{\frac {k_{{1}}h_{{2}}}{k_{{1}}h_{{2}}+h_{{1}}h_{{2}}+k_{{2}}h_{{1}}+k_{{1}}k_{{2}}}},{\frac {k_{{1}}k_{{2}
}}{k_{{1}}h_{{2}}+h_{{1}}h_{{2}}+k_{{2}}h_{{1}}+k_{{1}}k_{{2}}}}\right).
\end{array} \]
 
 \noi In the  case of the sum space we would obtain an average dynamics of  form:
 
 \[\begin{array}{ll}
 \displaystyle \dot{\bx}(t)=\theta_1\left[\frac{h_1}{k_1+h_1}\,A^{(1)}(\bx)+\frac{k_1}{k_1+h_1}\,A^{(2)}(\bx)\right]+\\[5mm]
\displaystyle \theta_2\left[\frac{h_2}{k_2+h_2}\,A^{(3)}(\bx)+\frac{k_2}{k_2+h_2}\,A^{(4)}(\bx)\right],
 \end{array}\]
 
\noi  with $\theta_1+\theta_2=1$. In the  case of the product space we would obtain an average dynamics of another form:
 
 \[\begin{array}{llll}
 \displaystyle \dot{\bx}(t)={\frac {h_{{1}}h_{{2}}}{k_{{1}}h_{{2}}+h_{{1}}h_{{2}}+k_{{2}}h_{{1}}+
k_{{1}}k_{{2}}}}\,A^{(1,3)}(\bx)+\\[5mm]
\displaystyle+{\frac {k_{{2}}h_{{1}}}{k_{{1}}h_{{2}}+h_{{1}}h_{{2}}
+k_{{2}}h_{{1}}+k_{{1}}k_{{2}}}}\,A^{(1,4)}(\bx)+\\[5mm]
\displaystyle+{\frac {k_{{1}}h_{{2}}}{k_{{1}}h_{{2}}+h_{{1}}h_{{2}}+k_{{2}}h_{{1}}+k_{{1}}k_{{2}}}}\,A^{(2,3)}(\bx)+\\[5mm]
+\displaystyle {\frac {k_{{1}}k_{{2}
}}{k_{{1}}h_{{2}}+h_{{1}}h_{{2}}+k_{{2}}h_{{1}}+k_{{1}}k_{{2}}}}\,A^{(2,4)}(\bx).
 \end{array}\]

 This example nicely shows that the direct sum space assumption leads to fractions of work done by the two macro-molecules, something which would become a splitting of the population working independently in different modes, based on the assumption there are fixed finite numbers of  identical macro-molecules working in each such mode. The product space creates dependencies among the states of the macro-molecules, creating a single mode of operation for the whole population.

\brem
Note that under suitable conditions the average dynamics may coincide. For example consider the case in which

\[A^{(1,3)}(\bx)=A^{(1,4)}(\bx)=A^{(1)}(\bx),~~~A^{(2,3)}(\bx)=A^{(2,4)}(\bx)=A^{(2)}(\bx).\]

\noi The average dynamics with $\Sig=\Sig_{prod}$ reduces to

\[\dot{\bx}(t)=\frac{h_1}{k_1+h_1}\,A^{(1)}(\bx)+\frac{k_1}{k_1+h_1}\,A^{(2)}(\bx), \]

\noi which is the average dynamics associated to the case $\Sig=\Sig_{sum}$ with $\theta_2=0$. The reason for this is that 
there exist general relations among the stationary measures associated to $\Sig_{sum}$ and $\Sig_{prod}$. One can easily check that
 for $\mu_{{sum},i}\neq 0$
  
\[\mu_{{sum},i}=\sum_{j}\mu_{{prod},(i,j)}.\]

\erem

\section{Applications and examples}

We now present important applications of the theory introduced above. The examples will illustrate how  the deterministic limit is essentially prescribed by the structure of the Markov chain, and the dynamics is affected by new nonlinear terms. Such new terms provide the effective dynamics 
originating from the average procedure. Note that in the examples the continuum approximation procedure will be recalled heuristically and in particular the continuum approximation of $\Kd^T$ can be computed according to  the theory developed in \cite{sbano1}.

\subsection{Enzyme kinetics}

This first, at the same time most important example, was already introduced as an illustrative example for the continuum and adiabatic limit in \cite{sbano1}. Now we finish it with the discussion of the determininistic limit. The Michaelis-Menten and the Hill's equation are often used to model reactions that exhibit a saturation behaviour. The classical derivation can be found in \cite{siegel}, and more examples and applications in \cite{keener}. Among other applications they constitute the basic set of enzyme-catalysed reactions, for example in metabolism of the cell (see \cite{siegel}). We show how an effect described by these kinetics can arise as a limit of a multiple-state system, where the switching takes place at high frequency. This will be an example of a system with one \idf~ and one \fdf. The system has state space $(m,O_i)\in\LL\times \Sig$ with $\Sig=\{O_0,O_1\}$ and let $\de_n,\tau_n$ be respectively the size and time scales. The \fdf~ $O_i$ is governed by the following set of reactions:

\begin{enumerate}

\item with rate $k^+(\de_n,\tau_n)/\tau_n$ and upon the binding of $1$ molecule of $A$ the state $O_0$  is transformed into the active state $O_1$

\[A+O_0 \rightarrow^{k^+(\de_n,\tau_n)/\tau_n} O_1,\]

\item with rate $k^-(\de_n,\tau_n)/\tau_n$ the state $O_1$ is decades into the inactive state $O_0$ releasing 
$A$ molecules

\[O_1 \rightarrow^{k^-(\de_n,\tau_n)/\tau_n} O_0+A.\]

\end{enumerate}


\noi We now prescribe the reactions taking place in each discrete state. In state $O_1$ a certain molecule $M$ is produced and degraded according to

\[\emptyset\rightarrow^{v(\de_n,\tau_n)/\tau_n}\,M\rightarrow^{\nu(\de_n,\tau_n)/\tau_n}\,\emptyset.\]

\noi Only in state $O_0$, the molecules $M$ are degraded  according to

\[M\rightarrow^{\nu(\de_n,\tau_n)/\tau_n}\,\emptyset.\]

\noi Obviously $M$ is an \idf., i.e. the number of small molecules can reach infinity. \\


\brem
We now assume $a>>1$. This implies 
\[P(t,m,a)=(P_0(t,m,a),P_1(t,m,a-\alpha))\simeq (P_0(t,m,a),P_1(t,m,a)).\]
\erem

\noi The master equation for the vector probabilities $P(t,m,a)$ can be written as

\[\frac{\pa P}{\pa t}=\Ld_n^*\,P+\Kd_n^T\,P, \]

where
\[
\Ld_n^*\, = \frac{1}{\tau_n}\left( \begin {array}{cc}
 \nu(\de_n,\tau_n)\,(\Eop^+-\id)(m\,\cdot \,) &0
 \\\noalign{\medskip}0 &  v(\de_n,\tau_n)\,(\Eop^--\id)(\,\cdot \,)+\nu(\de_n,\tau_n)\,(\Eop^+-\id)(m\,\cdot \,) \end {array} \right) ,
\]

with $\Eop^\pm P(m)=P(m\pm1)$, and

 \[
\Kd_n^T\, = \frac{1}{\tau_n}\, \left( \begin {array}{cc} 
a\,k^+(\de_n,\tau_n) & -k^-(\de_n,\tau_n)\\
\noalign{\medskip}-a\,k^+(\de_n,\tau_n) & k^-(\de_n,\tau_n)
\end {array} \right). 
\]

\noi As in section \ref{continuum} and in \cite{sbano1} it is shown that the continuum limit is

\[
\Lop^*\, = \, \left( \begin{array}{cc}
 \lop^0 & 0
 \\\noalign{\medskip}0 & \lop^1
 \end {array} \right) 
\]

 with

 \[
\lop^0=\frac{\partial}{\partial x}\,\nu\,x,~~\lop^1=\frac{\partial}{\partial x}\,(\nu\,x-v).
\]

and

\[
\Kd^T\, =  \left( \begin {array}{cc} 
a\,k^+ & -k^-
\\\noalign{\medskip}-a\,k^+& k^-
\end {array} \right).
\]

\noi The ME converges to
\[\frac{\pa \rho}{\pa t}=\Lop^*\,\rho+\frac{1}{\eps}\,\Kd\rho\]
with $\eps=O(\tau_n)$.\\
 From this we obtain the two vector fields

\[X_0(x)=-\nu\,x,~~~~X_1(x)=-\nu\,x+v.\]

\noi The matrix $\Kd^T$ has a unique invariant measure

\[\mu=\frac{1}{a\,k^++k^-}\left(\begin{array}{c}
k^-\\
a\,k^+
\end{array}\right).\]

\noi The \emph{average dynamics} is 

\[\frac{\pa f^{(0)}(x,t)}{\pa t}=\langle\be_\mu,\Lop^*(\mu(x)\,f^{(0)}(\bx,t))\rangle. \]

 After some algebra the equation for $f^{(0)}$ becomes

\begin{equation}
\frac{\pa f^{(0)}(x,t)}{\pa t}=\lop^0(\nu\,x\,f^{(0)}(x,t))
-\lop^1\left(\frac{v\,a\,k^+}{k^-+a\,k^+}f^{(0)}(x,t)\right).
\label{eq}
\end{equation}

\noi Equation (\ref{eq}) becomes

 \[\frac{\pa f^{(0)}(x,t)}{\pa t}=-\frac{\pa}{\pa x}\left(\left(-\nu\,x
+\frac{v\,a\,k^+}{k^-+a\,k^+}\right)f^{(0)}(x,t)\right), \]

\noi which is the Liouville equation. This equation is equivalent (see \cite{pavliotis}) to the time evolution of the concentration $x$ gouverned by the averaged vector field

\[X(x)=\frac{k^-}{a\,k^++k^-}\,X_0(x)+\frac{a\,k^+}{a\,k^++k^-}\,X_1(x). \]

\noi Hence the  average deterministic dynamics in this case  is

\begin{equation}
\frac{d x}{d t}=X(x)=
-\nu\,x+\frac{a\,k^+}{a\,k^++k^-}\,v.
\label{eq-alpha}
\end{equation}

\noi Here we see that the concentration $x$ of $M$ is produced at a rate that depends on the concentration $a$ of $A$,  with a saturation behaviour for $a$ large enough. 
 
 \brem
 Note that we can obtain a Michaelis-Menten kinetics.  Indeed let us consider the classical enzyme reaction
  
 \[A+E\rlha[k_{-1}]{k_1} C\ra^{k_2} X+E. \]
 
 \noi The time scale analysis leads to the Michaelis-Menten rate equation
 
 \begin{equation}
 \frac{d x(t)}{dt}=\frac{k_2\,e_0\,a}{a+(k_{-1}/k_{1})},
 \label{MM}
 \end{equation}
  
where $e_0$ is the steady state for the enzyme concentration $[E]$. Taking $\de=0$ and $\alpha=1$ in (\ref{eq-alpha}) we recover (\ref{MM}) by setting

\[v=k_2v\mbox{ and } k_{-1}/k_{1}=k^-/k^+.\]
 \erem
 
For the two-state system we can also construct the noise. In fact the FPE associated to the marginal distribution $f$ is

 \[\frac{\pa f}{ \pa t}=\langle\be_\mu,\Lop^*(\mu\,f)\rangle-\eps\langle\be_\mu,(\Lop^*\,(K_\mu^T)^D\Lop^*)(\mu\,f)\rangle, \]
 
 and 
 
 \[f(t,x)=\rho_0(t,x)+\rho_1(t,x). \]
 
\noi The FPE is given by
 
 \begin{eqnarray*}
 \frac{\pa f(x,t)}{\pa t}=\frac{\partial}{\partial x}\left[\left(
-\nu\,x+\frac{a\,k^+}{a\,k^++k^-}\,v\right)f(x,t)\right]+\\
+\eps\,{\frac {a{\it k^+}\,{\it k^-}}{ \left( {\it k^-}+a\,{\it k^+} \right) ^{3}}}\, \left[\left(\frac{\partial}{\partial x}\,(2\nu\,x-v)\,\left(\frac{\partial}{\partial x}\,(2\nu\,x-v)f(x,t)\right)\right)\right].
\end{eqnarray*}

\noi From the FPE one can then derive the associated SDE

\[\begin{array}{ll}
\displaystyle dx(t)= \left(
-\nu\,x(t)+\frac{a\,k^+}{a\,k^++k^-}\,v-\eps\frac{a\,k^+\,k^-}{a\,k^++k^-}\,(2\nu\,x(t)-v)\right)\,dt \\[5mm]
\displaystyle + \sqrt{\veps\,{\frac {a{\it k^+}\,{\it k^-}}{ \left( {\it k^-}+a\,{\it k^+} \right) ^{3}}}\,(2\nu\,x(t)-v)^2+
\frac{\de}{2}\,\left(\nu\,x(t)+\frac{a\,k^+}{a\,k^++k^-}\,v\right)}dw_t.
\end{array} \]

\subsection{Formation of macromolecules}

We now analyse the formation of a large macro-molecule like a protein formed at the ribosomes, using the mRNS as a matrix. But for simplicity we will not distinguish between different types of amino acids. Such a process is often modelled by using a generalisation of the Hill's kinetics. The process takes place in several steps, namely each new molecule is formed after a sequence of reactions is completed. This sequentiality introduces a cooperative behaviour. We assume that the process occurs in $g$ steps. The $E$ molecules  react with a substrate $S$ in $g$ consecutive reactions,  and only after the last reaction is terminated a molecule of $P$ is formed. Such a process can be described by the following chain of reactions:
 
 \[E\rlha[\nu]{k\,S}ES\rlha[\nu]{k\,S}ES_2\rlha[\nu]{k\,S}ES_3\rlha[\nu]{k\,S}...\rlha[\nu]{k\,S}ES_g\rightarrow^{k_p} P \]
 
\noi We obtain a system with a state space 

\[\Sigma=\N\times\{E,ES,ES_2,...,ES_g\}. \]

\noi Let $P(p,t)=(P_0(p,t),...,P_g(p,t))$.  The master equation is given by

\begin{equation}
\left\{\begin{array}{lllll}
\dot{P}_0(p,t)=-k\,s\,P_0(p,t)+\nu\,P_1(p,t)\\
\dot{P}_1(p,t)=k\,s\,P_0(p,t)-\nu\,P_1(p,t)-k\,(s-1)\,P_1(p,t)+\nu\,P_2(p,t)\\
\dot{P}_2(p,t)=k\,(s-1)\,P_1(p,t)-\nu\,P_2(p,t)-k\,(s-2)\,P_2(p,t)+\nu\,P_3(p,t)\\
\ldots\\
\dot{P}_g(p,t)=k\,(s-g+1)\,P_{g-1}(p,t)-\nu\,P_g(p,t)-k_p\,P_g(p,t)+k_p\,P_{g}(p-1,t).\\
\end{array}\right.
\label{buildingup}
\end{equation}

\noi Let $s$ be the number of particles of type $S$. Let us assume TO BE CHANGED

\[k=\frac{k_0}{\eps},~~\nu=\frac{\nu_0}{\eps}\mbox{, with $\eps$ small.}\]

\noi For simplicity we consider $g=3$. Then this equation can rewritten as follows

\[\frac{\pa P}{\pa t}=\Ld^*\, P+\frac{1}{\eps}\Kd^T(s)\,P, \]

where 

\[
\Ld^*\, = \frac{1}{\tau}\, \left( \begin {array}{ccccc}
 0&0 & 0 & 0\\
 0 & 0&0&0\\
 0&0&0&0\\
 0&0&0&k_p\,(\Eop^--\id)(\,\cdot \,)
  \end {array} \right), 
\]

 and

 \[
\Kd^T(s)\, = \frac{1}{\tau}\, \left( \begin {array}{ccccc}
 -k\,s &\nu & 0 & 0\\
 k\,s & -\nu-k\,(s-1)&\nu&0\\
 0&k\,(s-1)&-\nu-k\,(s-2)&\nu\\
 0&0&k\,s&-\nu
  \end {array} \right).
\]

\noi In the continuum approximation we consider $s$ as a parameter.
For every $s$ the matrix $K(s)$ is a generator of an ergodic Markov chain. Indeed $K^T(s)$ has generically only one zero eigenvalue. The eigenvalues are given by the zeros of 

\[\begin{array}{lll}
\det(\Kd-zI)=
{z}^{4}+ \left( 3\,k(s-1)+3\,\nu\right) {z}^{3}+\\[2mm]
+ \left( 4\,k\nu(s-1)+3\,{k}^{2}{s}(s-2)+3\,\nu^{2}+\,k\nu+2\,{k}^{2} \right) {z}^{2}+\\[3mm]
+ \left( ks{\nu}^{2}+{k}^{2}\nu\,s(s-1)+2\,k^3\,s+{k}^{3}s^2(s-3)+{\nu}^{3} \right) z=0.
\end{array}\]

\noi For $s\ge g=3$ there is only one zero eigenvalues and the others are 
strictly negative. The MC is ergodic and its unique invariant measure is equal to

\[
\mu(s)=\frac{1}{\nu_0^3+k_0\nu_0^2k_0s+\nu_0k_0^2s(s-1)+k_0^3s(s-1)(s-2)}\,\left(\begin{array}{c}
\nu_0^{3}\\[2mm]
{\nu_0}^{2}k_{0}s\\[2mm]
\nu_0\,{k_{0}}^{2}{s}(s-1)\\[2mm]
 {k_{0}}^{3}s(s-1)(s-2)
\end{array}
\right).\]

\noi From the form of $\Ld^*$ we construct its continuum approximation $\Lop^*$. Taking only first order terms we obtain

\[
\Lop^*\, = \, \left( \begin {array}{ccccc}
 0&0 & 0 & 0\\
 0 & 0&0&0\\
 0&0&0&0\\
 0&0&0&-k_p\frac{\pa}{\pa x_p}
  \end {array} \right) 
\]

\noi Therefore the deterministic equation for the concentration $x_p$ of $P$ is

\[\frac{d x_p}{dt}=\mu_4(s)k_p=
\frac{ {k}^{3}s(s-1)(s-2)}{\nu^3+k\nu^2s+\nu\,k^2s(s-1)+k^3s(s-1)(s-2)}\,k_p, \]

\noi where $s$ again  is the number of particles  of the substrate.
The last equation gives the precise expected concentration of the product, i.e. formed macro-molecules like a protein, where $s$ in this case would be modelling the number of individual amino-acids ready for assembly in each step to be attached to the polymer.\\
 For large $s$ the kinetics reads 
\[\frac{d x_p}{dt}=\mu_4(s)k_p=
\frac{ {k}^{3}\,s^3}{\nu^3+k\nu^2s+\nu\,k^2\,s^2+k^3\,s^3}\,k_p. \]
which is a generalised Hill's term.

\brem
Note that in general the number of states $g$ is the maximal exponent in the rational function 
which gives the effective reaction rate.
\erem

\subsection{Averaging the average}

This example is meant to illustrate the consequences of having processes which 
are independent but interact  through a common background. We have already discussed the action of several identical enzymes present in a cell as the typical application. Let us consider the following set of reactions

\[\begin{array}{lll} 
O_1\rlha[h]{k\,A_1}O_2\rightarrow^{\alpha}O_2+X\\[4mm]
O_4\rlha[h]{k\,A_2}O_3\rightarrow^{\alpha}O_3+X\\[4mm]
X\ra^{\gamma}\emptyset.
\end{array}\]

\noindent Here we have in principle two MCs: $\Sig_1=\{O_1,O_2\}$ and $\Sig_2=\{O_3,O_4\}$. Let us consider the direct sum of them which corresponds to $4$-state space

\[S=\{O_1,O_2,O_3,O_4\}. \]

\noi Each $O_i$  is interacting with a substrate $A$  in two distinguished pools, $A_1$ and $A_2$, with number of particles  $a_1$ and $a_2$, respectively. We shall assume that $A_1$ and $A_2$ remain discrete and therefore don't contribute to the continuum approximation. For brevity we skip the construction of the ME and give only the FPE

\[\frac{\pa \rho(x,t)}{\pa t}=\Lop^*(\rho(x,t))+\frac{1}{\eps}\Kd^T(x)\rho(x,t), \]

where

\[
\Lop^*\, = \, \left( \begin {array}{ccccc}
 \frac{\pa}{\pa x}(\gamma\cdot)&0 & 0 & 0\\
 0 & \frac{\pa}{\pa x}(\gamma\cdot)-\frac{\pa}{\pa x}(\alpha\cdot)&0&0\\
 0&0&\frac{\pa}{\pa x}(\gamma\cdot)&0\\
 0&0&0&\frac{\pa}{\pa x}(\gamma\cdot)-\frac{\pa}{\pa x}(\alpha\cdot)
  \end {array} \right) 
\]

 and

 \[
\Kd^T(x)\, = \, \left( \begin {array}{ccccc}
 -a_1k &h & 0 & 0\\
 a_1k & -h&0&0\\
 0&0&-a_2k&h\\
 0&0&a_2k& -h
  \end {array} \right). 
\]

\noindent We can easily see that in this case $\dim(M_K)=2$. In fact the two stationary measures are given by

\[
\mu^{(1)}=\frac{1}{a_1k+h}\,\left(\begin{array}{c}
h\\[2mm]
a_1k\\[2mm]
0\\[2mm]
0 
 \end{array}\right),~~~
\mu^{(2)}=\frac{1}{a_2k+h}\,\left(\begin{array}{c}
0\\[2mm]
0\\[2mm]
h\\[2mm]
a_2k
\end{array}
\right).\]

\noindent We take the convex combination

\[\mu=\theta_1\mu^{(1)}+\theta_2\mu^{(2)}, \]

\noindent with $\theta_1+\theta_2=1$, and construct the the FPE in the limit $\eps=0$. This is the deterministic limit 
and the FPE becomes the Liouville equation

\[\frac{\pa f}{\pa t}=\langle\be_\mu,\Lop^*(\mu\,f)\rangle.\]

\noindent It turns out that the right hand side is

\[
\langle\be_\mu,\Lop^*(\mu\,f(x,t))\rangle=2\frac{\pa (\gamma x f(x,t))}{\pa x}+
\frac{\pa}{\pa x}\left[-\theta_1\,\frac{\alpha\,k\,a_1}{ka_1+h}-\theta_2\,\frac{\alpha\,k\,a_2}{ka_2+h}\right]f(x,t).\]

\noindent Hence the average dynamics for the concentration of $X$ has the following form

\begin{equation}
\frac{d x(t)}{dt}=-2\,\gamma\,x(t)+ \alpha\,k\,\left[\theta_1\,\frac{a_1}{ka_1+h}+\theta_2\,\frac{a_2}{ka_2+h}\right] \mbox{ with $\theta_1+\theta_2=1$.}
\label{averages}
\end{equation}

\noindent By inspection of (\ref{averages}) one can understand the motivation to call  
 this section "averaging the average". In fact, the average dynamics results from 
averaging over $\mu$, this is the first average. Now $\mu$ is a convex combination. One variable  (in this case $x$) is affected by the sub-chains $\Sig_1,\Sig_2$, hence the deterministic dynamics contains terms depending on both convex parameters. In particular, in this case we have the term

\begin{equation}
\label{average2}
\theta_1\,\frac{a_1}{ka_1+h}+\theta_2\,\frac{a_2}{ka_2+h},
\end{equation}

\noindent where the convex parameters are $\theta_1$ and $\theta_2$. 
Due to the normalisation of the convex parameters, terms like (\ref{average2})
 can be interpreted as averages, namely they describe how a variable influenced by the states of different closed sub-chains is affected in the adiabatic approximation.

\subsubsection{Many independent molecular machines}

The previous system permits us to make a further generalisation, leading closer to a finite population of molecular machines like a larger but finite number of enzymes in a reaction volume. Let us assume to have $M$ such molecular mechanisms or machines whose dynamics is specified by the following reactions

\[ A_i+O_i^-\rlha[h_i]{k_i}O_i^+, ~~~i=1,...,M, \]

\noi each of which produces $X$ according to the scheme

\[O_i^+\ra^{\gamma_i}O_i^++X.\]

\noi Furthermore we let $X$ degrade, i.e.

\[X\ra^{\de}\emptyset.\]

\noi Each machine has a discrete space $\Sig_i=\{O_i^-,O_i^+\}$ and has a MC with generators $\Kd_i$ such that

\[\Kd_i^T=\left(\begin{array}{cc}
-k_i\,a_i & h_i\\
k_i\,a_i & -h_i
\end{array}\right).\]

\noi The total space is the direct sum of the $S_i$'s, namely the collection

\[\Sig=\{O_1^{-},O_1^{+}....,O_M^{-},O_M^{+}\}. \]

\noi Note that there are no processes linking the state of the $i$-th machine to the state of the $j$-th machine. This implies that we have a direct product of MCs whose generator $\Kd$ has a diagonal form

\[\Kd=\left(\begin{array}{cccc}
\Kd_1 & 0&\dots & 0\\
0 &  \Kd_2& \ldots &0\\
\vdots & \ldots & \dots &\vdots\\
0&0&\ldots &\Kd_M
\end{array}\right).\]

\noi One can easily verify that the invariant measure of $\Kd^T$ is given by

\[\mu=(\mu_1(a_1),...,\mu_M(a_M)),\]

\noi where each $\mu_i(a_i)$ is a two-dimensional vector which is an invariant measure for $\Kd_i^T$. Explicitly we have

\[\mu_i(a_i)=\left(\frac{k_i\,a_i}{k_i\,a_i+h_i},\frac{k_i\,a_i}{k_i\,a_i+h_i}\right).\]

\noi Now let us fix a convex combination of measures

\begin{equation}
\label{mu}
\mu(a_1,...a_M)=\sum_{i=1}^M\mu_i(a_i)\,\theta_i,
\end{equation}

\noi where the average dynamics for $x$ will be given by

\[\frac{d x(t)}{d t}=-\de\,x(t)+\sum_{i=1}^M\gamma_i\,\theta_i\,\mu_i(a_i).\]

\brem
Let us now consider the normalisation condition on $\theta_i$'s.  If we set
\begin{equation}
\label{norm1}
\sum_{i=1}^M\theta_i=1
\end{equation}
then the $k$th component of $\mu$ is interpreted as the probability that 
one machine is one of the $2M$ $k$th-state  of $\Sigma$. It can be useful 
to consider another normalisation given by
\begin{equation}
\label{normM}
\sum_{i=1}^M\theta_i=M.
\end{equation}
In this case the $k$th component of $\mu$ is interpreted as the number of 
machines in the $k$th state of a $2\times 2$ MC.
\erem

\noi Let us now fix normalisation (\ref{normM}) and suppose that all the $M$ machines are equal. This implies

\[\gamma_i=\gamma,~~a_i=a\mbox{, for all $1 \le i \le M$}. \]
\noi For the average dynamics this implies 

\[\frac{d x(t)}{d t}=-\de\,x(t)+ M \gamma\mu(a),\]

\noi which allows us to conclude that $M$ completely  independent molecular machines lead to 
an average dynamics which expression is equivalent to the one of a single machine. Only the rate of conversion, as expected, scales with the number of enzymes present in the system.

\subsubsection{Different fractions of independent molecular machines}
The preceding examples are useful to construct the following generalisation. Consider $M$ molecular 
machines  defined by 
 
\[ O_{i}\rlha[h_{i}]{k_{i}\,A_{i}}O_{i+2}\rightarrow^{\alpha_i}O_{i+2}+X, \]

with $i=1,...,M-2$ and degradation

\[X\ra^\de\emptyset.\]

 \noi  Using the construction explained in the previous paragraphs one shows that 
 there are $M$ Markov chains, with state spaces $\Sig_{i}=(O_{i},O_{i+2})$, total space $\Sig=\oplus_{i=1}^M\Sig_i$, and with invariant measures
  
  \[\mu_i(a_i)=\left(0,...,0,\frac{k_i\,a_i}{k_i\,a_i+h_i}, \frac{h_i}{k_i\,a_i+h_i},0,...,0\right).\]  

\noi From the invariant measures one can easily obtain the average dynamics

\[\dot{x}(t)=-\de x(t)+\sum_{i=1}^M\theta_i\,\alpha_i\,\frac{k_i\,a_i}{k_i\,a_i+h_i}, \]

where we use normalisation (\ref{normM}). Now if the $M$ molecular machines are subdivided into 
$Q$ classes of equal machines, then $\theta_i$s can be interpreted as the fractions
 of active machines. Therefore 
 
 \[\theta_i=M_i, \]

and the average dynamics reads

\[\dot{x}(t)=-\de x(t)+\sum_{i=1}^Q\theta_i\,\alpha_i\,\frac{k_i\,a_i}{k_i\,a_i+h_i}.\]

\noi In such a case the average dynamics is equivalent to having $Q$ types of molecular machines,  each of which contributes with a new rate $\alpha_i'$ equal to

\[\alpha_i'=\alpha_i\,\theta_i=\alpha_i\,M_i,\mbox{ $i=1,...,Q$}. \]

\subsection{State changes according to a discrete transport model}

\begin{figure}[htbp] 
   \centering
   \includegraphics[width=10cm]{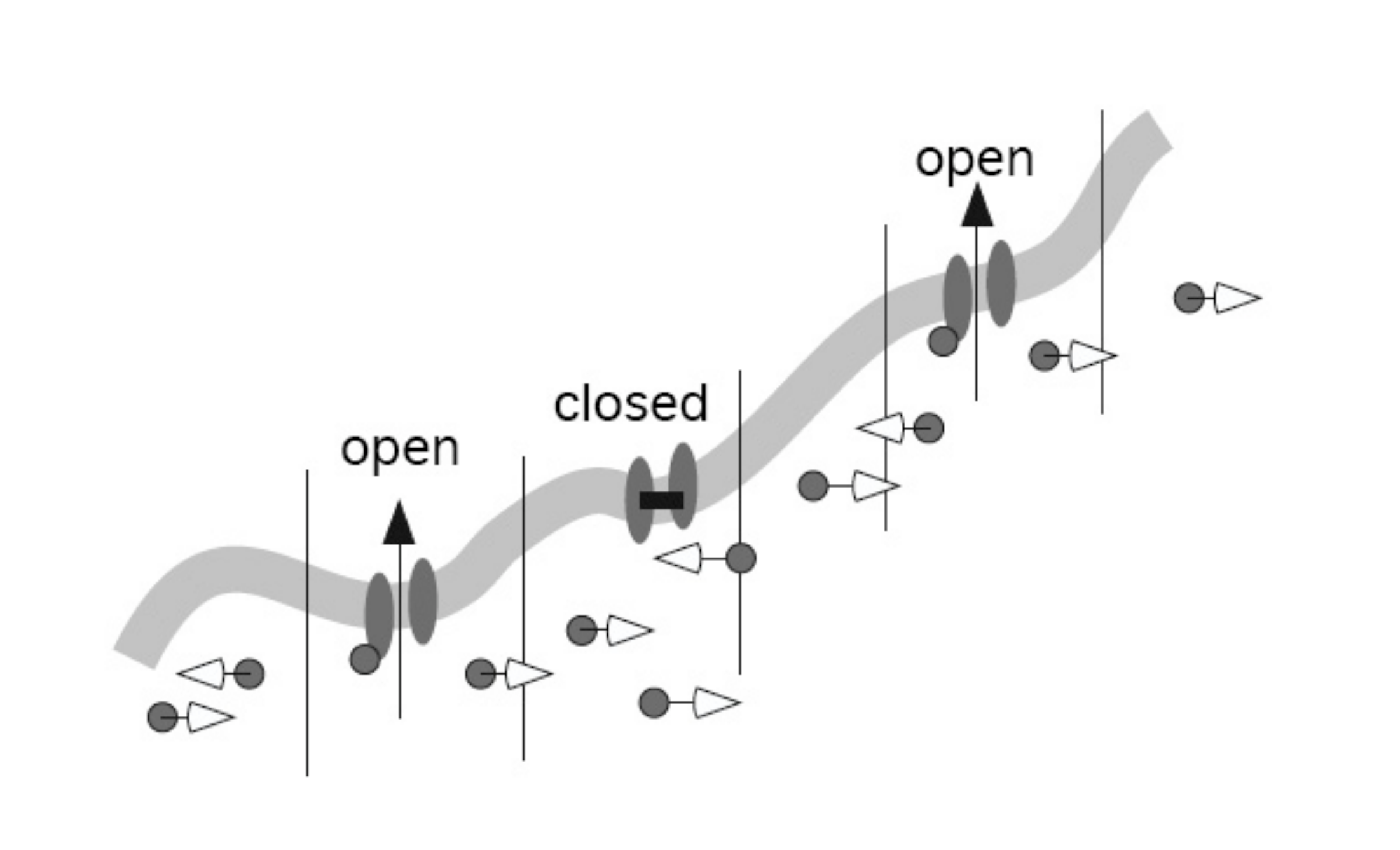} 
   \caption{ \small The discrete transport model interpreted to describe single ion channels that can be either open or closed, depending on whether a signalling molecule forms or does not form a complex with the channel. The space is discretised along the membrane in one dimension. The signalling molecules diffuse and therefore either enter or leave any spatial compartment (most likely by diffusion from which the transition rates $\alpha_i$ and $\beta_i$ would have to be computed), and this is indicated by the vectors attached to each signalling molecule.}
   \label{fig:membrane}
\end{figure}

We next like to consider a system formed by $N$ compartments or spatial locations, where 
a certain molecule $A$ can pass from one compartment/location  to the other.  Further we assume that from some of these compartments a molecule $X$ is produced or transported upon the activation 
of a molecular machinery. This is a sketch of two typical examples. Either the macro-molecule models a simple channel, a membrane protein which opens or closes the membrane for a molecule of type $X$, according to the availability of the molecule of type $A$. Or the system can be interpreted as a sketch of a genetic system, where genes are switched on to produce mRNA (type $X$ in this case), according to the availability of the transcription factor, in this interpretation type $A$. One interpretation is illustrated in Figure \ref{fig:membrane}.

We denote by $A_i$ the $A$ molecules in the $i$-th compartment. Consider the following chain of reactions   

 \[\emptyset\ra^{\alpha_0}A_1\rlha[\beta_2]{\alpha_1}A_2\rlha[\beta_3]{\alpha_2}A_3\rlha[\beta_4]{\alpha_3}...\rlha[\beta_{N}]{\alpha_{N-1}}A_N\]

\noi These reactions describe a discrete approximation of $A$ being transported in the $N$
 compartments. We assume that there is a subset $L\subset [1,...,N]$ of compartments at which particles can trigger the molecular machinery. Let $|L|=l\leq N$. Let $i\in L$, a molecule of $A_i$  can bind to the site (an operator or channel) $O_i$ according to the following reactions

\[A_i+O_-^i\rlha[k_i]{h_i}O^i_+.\]

\brem
As we have already seen in other examples, there is a Markov Chain with generator
\[\Kd_i=\left(\begin{array}{cc}
-k_ia_i & h_i\\
k_ia_i & -h_i
\end{array}\right)\]

\noi for the two states $(O_i^-,O_i^+)$. We also have the invariant measure

\[\mu_i=\left(\frac{k_ia_i}{k_ia_i+h},\frac{h_i}{k_ia_i+h}\right), \]

\noi where as usual $[A_i]=a_i.$
\erem

\noi We now assume that a molecule $X$ is produced/transported through the $i$th 
machinery

\[O^i_+\ra^{\gamma_i} O^i_++X.\]

\noi  To close the system we include that  molecules degrade according to

\[A_i\ra^{\de_A}\emptyset,~~~X\ra^{\de_X}\emptyset.\]

\noi The operators $O_i$ form a MC whose states are $g=2^l$ and 

\[\Sig=\{\sig=(O_{i_1},...,O_{i_l})\mbox{ with } O_i\in \{O_i^-,O_i^+ \}\}.\]

\noi Now the processes in which $X$ is produced/transported read

\[(O_{i_1},...,O_k^+,...,O_{i_l})\ra^{\gamma_k}(O_{i_1},...,O_k^+,...,O_{i_l})+X
\mbox{, for $k=i_1,...,i_l$.}\]

\noi For the sake of simplicity let $L=\{p,q\}$, we have $g=4$ and 

\[\Sig=\{(O_p^-,O_q^-),(O_p^+,O_q^-),(O_p^+,O_q^+),(O_p^-,O_p^+)\}, \]

\noi with
\begin{equation}
\begin{array}{lll}
(O_p^+,O_q^-)\ra^{\gamma_p}(O_p^+,O_q^-)+X, \\[4mm]
(O_p^+,O_q^+)\ra^{\gamma_p}(O_p^+,O_q^+)+X, \\[4mm]
(O_p^+,O_q^+)\ra^{\gamma_q}(O_p^+,O_q^+)+X, \\[4mm]
(O_p^-,O_q^+)\ra^{\gamma_q}(O_p^-,O_q^+)+X.
\end{array}
\label{noninteracting}
\end{equation}

\noi The associated matrix $\Kd^T$ is

\[
\Kd^T=\left( \begin {array}{cccc} -k_{{p}}a_p-k_{{q}}a_q&h_{{p}}&h_{{q}}&0
\\\noalign{\medskip}k_{{p}}a_p&-h_{{p}}-k_{{q}}a_q&0&h_{{q}}
\\\noalign{\medskip}k_{{q}}a_q&0&-h_{{q}}-k_{{p}}a_p&h_{{p}}
\\\noalign{\medskip}0&k_{{q}}a_q&k_{{p}}a_p&-h_{{q}}-h_{{p}}\end {array}
 \right).\]

\noi Its invariant measure is
\[
\begin{array}{ll}
\displaystyle\mu(a_p,a_q)=\left({\frac {h_{{p}}h_{{q}}}{h_{{p}}h_{{q}}+h_{{q}}k_{{p}}a_{{p}}+k_{{q}}a
_{{q}}k_{{p}}a_{{p}}+k_{{q}}a_{{q}}h_{{p}}}},{\frac {h_{{q}}k_{{p}}a_{
{p}}}{h_{{p}}h_{{q}}+h_{{q}}k_{{p}}a_{{p}}+k_{{q}}a_{{q}}k_{{p}}a_{{p}
}+k_{{q}}a_{{q}}h_{{p}}}},\right. \\[4mm]
\displaystyle \left.{\frac {k_{{q}}a_{{q}}k_{{p}}a_{{p}}}{h_{{p}}h_{{q}}+h_{{q}}k_{{p}}a_{{p}}+k_{{q}}a_{{q}}k_{{p}}a_{{p}}+k_{{q}}a_{{q}}h_{{p}}}},{\frac {k_{{q}}a_{{q}}h_{{p}}}{h_{{p}}h_{{q}}+h_{{q}}k_{{p}}a_{{p}}+k_{{q}}a_{{q}}k_{{p}}a_{{p}}+k_{{q}}a_{{q}}h_{{p}}}}
\right).
\end{array} \]

\noi In general $K^T$ will be more complicated but still ergodic. In fact any state in $S$ can be reached from any other state. The MC has a unique invariant measure $\mu(a_L)$ where $a_L=(a_{i_1},...a_{i_l})$. The deterministic dynamics in terms of the concentrations $a_i$ and $[X]=x$
 is given by

\begin{equation}
\left\{\begin{array}{llllll}
\dot{a}_1(t)=\alpha_0+\beta_2a_2(t)-(\alpha_1+\de_A)a_1(t)\\
\dot{a}_i(t)=\alpha_{i-1}a_{i-1}(t)+\beta_{i+1}a_{i+1}(t)-(\alpha_i+\beta_i+\de_A)a_i(t),~~i=2,...,N-1\\
\ldots\\
\dot{a}_N(t)=\alpha_{N-1}a_{N-1}(t)-(\beta_N+\de_A)a_N(t)\\
\dot{x}(t)=-\de_X\, x(t)\\
\dot{x}(t)=-\de_X\, x(t)+\gamma_s\mbox{ for $s\in S$}
\end{array}\right.
\label{AX-dynamics}
\end{equation}

\noi The average dynamics will therefore be

\begin{equation}
\left\{\begin{array}{llll}
\dot{a}_1(t)=\alpha_0+\beta_2a_2(t)-(\alpha_1+\de_A)a_1(t)\\
\dot{a}_i(t)=\alpha_{i-1}a_{i-1}(t)+\beta_{i+1}a_{i+1}(t)-(\alpha_i+\beta_i+\de_A)a_i(t),~~i=2,...,N-1\\
\ldots\\
\dot{a}_N(t)=\alpha_{N-1}a_{N-1}(t)-(\beta_N+\de_A)a_N(t)\\[3mm]
\dot{x}(t)=-\de_X\, x(t)+\sum_{s\in S}\gamma_s\mu_s(a_L)
\end{array}\right.
\label{AX-average-dynamics}
\end{equation}

\noi For the simple case $L=\{p,q\}$ we have:

\begin{equation}
\left\{\begin{array}{lllll}
\dot{a}_1(t)=\alpha_0+\beta_2a_2(t)-(\alpha_1+\de_A)a_1(t)\\
\dot{a}_i(t)=\alpha_{i-1}a_{i-1}(t)+\beta_{i+1}a_{i+1}(t)-(\alpha_i+\beta_i+\de_A)a_i(t),~~i=2,...,N-1\\
\ldots\\
\dot{a}_N(t)=\alpha_{N-1}a_{N-1}(t)-(\beta_N+\de_A)a_N(t)\\[3mm]
\dot{x}(t)=-\de_X\, x(t)+\gamma_p(\mu_2(a_p(t),a_q(t))+\mu_3(a_p(t),a_q(t)))+\\[3mm]
+\gamma_q(\mu_1(a_p(t),a_q(t))+\mu_4(a_p(t),a_q(t)))
\end{array}\right.
\label{l=2}
\end{equation}

\noi Now note that

\[\begin{array}{ll}
\displaystyle \mu_2(a_p,a_q)+\mu_3(a_p,a_q)=\frac{k_pa_p}{h_p+k_pa_p}, \\[4mm]
\displaystyle \mu_1(a_p,a_q)+\mu_4(a_p,a_q)=\frac{k_qa_q}{h_q+k_qa_q}.
 \end{array}\]

 \noi Therefore the equation for $x$ in turn becomes

\[\dot{x}(t)=-\de_X\, x(t)+\gamma_p\,\frac{k_pa_p(t)}{h_p+k_pa_p(t)}+
\gamma_q\,\frac{k_qa_q(t)}{h_q+k_qa_q(t)}.\]

\brem
Since the MC is a product of two $2$-states MCs, 
we have that summing the invariant measure $\mu$ over the possible states of one MC  produces 
the component of the invariant measure of the other chain. The form of the reactions 
 (\ref{noninteracting}) implies that the contribution of the compartments
  $A_p$ and $A_q$ to the dynamics of $x$ are \emph{uncoupled}.
\erem 

\brem
Consider a modification of reactions (\ref{noninteracting}) into

\begin{equation}
\begin{array}{lll}
(O_p^+,O_q^-)\ra^{\gamma_1}(O_p^+,O_q^-)+X, \\[4mm]
(O_p^+,O_q^+)\ra^{\gamma_2}(O_p^+,O_q^+)+X, \\[4mm]
(O_p^+,O_q^+)\ra^{\gamma_3}(O_p^+,O_q^+)+X, \\[4mm]
(O_p^-,O_q^+)\ra^{\gamma_4}(O_p^-,O_q^+)+X,
\end{array}
\label{interacting}
\end{equation}

\noi where $\gamma_i\neq\gamma_k$ for $i\neq k$. This would imply 
that the production/transport of $X$ \emph{always} 
depends on both states $O_p$ and $O_q$. Even though $A_p$ and $A_q$ are 
far a part in the chain of the compartments their contributions 
to the dynamics of $X$ are coupled. Indeed one easily find that in this case 
the dynamics of $X$ is

\[\begin{array}{ll}
\dot{x}(t)=-\de_X\, x(t)+\gamma_1\mu_1(a_p(t),a_q(t))
+\gamma_2\mu_2(a_p(t),a_q(t))+\\[3mm]
+\gamma_3\mu_3(a_p(t),a_q(t))+\gamma_4\mu_4(a_p(t),a_q(t))
\end{array}
\]

\noi which cannot be reduced to an expression of the form

\[\dot{x}(t)=-\de_X\, x(t)+f(a_p(t))+g(a_q(t))\]
 
 \noi for some smooth functions $f,g$.

\erem

\section{Discussion}

We have presented a rational and mathematically sound  derivation of models describing the non-spatial dynamics of large macro-molecular machines finite in number that interact with smaller, 'communicating', 'signalling' or 'substrate'-forming molecules in the cell. The approach can be used to test various assumptions in one framework, like investigating the effect of small numbers of such molecules on the performance  of the larger machines, or to recover different types of enzyme kinetics by considering the deterministic limit only. Here the main advantage is that the microscopic assumptions can be clearly stated, allowing the framework presented in this series of papers to serve as a tool for model construction. The main directions to be discussed further should be a more systematic investigation of the noise expected in such molecular systems when the smaller molecules vary heavily in numbers. An interesting extension is to analyse the problem of large deviations in order to describe situations of bi-stability in the macroscopic equations, again under the influence of noise. This requires to incorporate the stochastic fluctuations (e.g. small occupation numbers effect and noise) on long time scales. Yet another  direction is to consider the adiabatic theory in the context of a many-body approach, see for example \cite{doi1} and \cite{sasai-wolynes}.




\end{document}